\documentclass[journal=nalefd,manuscript=letter]{achemso}

\usepackage{graphicx}
\usepackage{amssymb}
\usepackage{amsmath}
\usepackage{bm}
\usepackage{color}
\usepackage[breaklinks]{hyperref}
\usepackage[all]{hypcap}

\hypersetup{
    plainpages=false,
    bookmarks=false,         
    unicode=false,          
    pdfborder={0 0 0}
    pdftoolbar=true,        
    pdfmenubar=true,        
    pdffitwindow=false,     
    pdfstartview={FitH},    
    pdftitle={},    
    pdfauthor={},     
    pdfproducer={LNCMI-CNRS-UGA-UPS-INSA}, 
    pdfkeywords={High} {Magnetic} {Fields}, 
    pdfnewwindow=true,      
    linktoc=section,
    colorlinks=true,       
    linkcolor=blue,          
    citecolor=red,        
    filecolor=magenta,      
    urlcolor=blue           
}

\author{Z. Yang}
\affiliation{Laboratoire National des Champs Magn\'etiques Intenses,
CNRS-UGA-UPS-INSA, 143, avenue de Rangueil, 31400 Toulouse}

\author{A. Surrente}
 \affiliation{Laboratoire National des Champs Magn\'etiques Intenses, CNRS-UGA-UPS-INSA, 143, avenue de Rangueil, 31400 Toulouse}

\author{G. Tutuncuoglu}
\affiliation{Laboratory of Semiconductor Material, Ecole
Polytechnique F\'ed\'erale de Lausanne, Lausanne, Switzerland}

\author{K. Galkowski}
\affiliation{Laboratoire National des Champs Magn\'etiques Intenses,
CNRS-UGA-UPS-INSA, 143, avenue de Rangueil, 31400 Toulouse}
\altaffiliation{Institute of Experimental Physics, Faculty of
Physics, University of Warsaw - Pasteura 5, 02-093 Warsaw, Poland}

\author{M. Cazaban-Carraz\'e }
 \affiliation{Laboratoire National des Champs Magn\'etiques Intenses, CNRS-UGA-UPS-INSA, 143, avenue de Rangueil, 31400 Toulouse}

\author{F. Amaduzzi}
\affiliation{Laboratory of Semiconductor Material, Ecole
Polytechnique F\'ed\'erale de Lausanne, Lausanne, Switzerland}

\author{P. Leroux}
\affiliation{Laboratory of Semiconductor Material, \'Ecole
Polytechnique F\'ed\'erale de Lausanne, Lausanne, Switzerland}

\author{D. K. Maude} \affiliation{Laboratoire National des Champs Magn\'etiques Intenses, CNRS-UGA-UPS-INSA, 143, avenue de Rangueil, 31400 Toulouse}

\author{A. Fontcuberta i Morral}
 \email{anna.fontcuberta-morral@epfl.ch}\affiliation{Laboratory of Semiconductor Material, Ecole
Polytechnique F\'ed\'erale de Lausanne, Lausanne, Switzerland}

\author{P. Plochocka}
 \email{paulina.plochocka@lncmi.cnrs.fr}\affiliation{Laboratoire National des Champs Magn\'etiques Intenses,
CNRS-UGA-UPS-INSA, 143, avenue de Rangueil, 31400 Toulouse}

\title[]
{Revealing large-scale homogeneity and trace impurity sensitivity of
GaAs nanoscale membranes}

\begin{document}

\begin{abstract}
III-V nanostructures have the potential to revolutionize
optoelectronics and energy harvesting. For this to become a reality,
critical issues such as reproducibility and sensitivity to defects
should be resolved. By discussing the optical properties of MBE
grown GaAs nanomembranes we highlight several features that bring
them closer to large scale applications. Uncapped membranes exhibit
a very high optical quality, expressed by extremely narrow neutral
exciton emission, allowing the resolution of the more complex
excitonic structure for the first time. Capping of the membranes
with an AlGaAs shell results in a strong increase of emission
intensity but also to a shift and broadening of the exciton peak.
This is attributed to the existence of impurities in the shell,
beyond MBE-grade quality, showing the high sensitivity of these
structures to the presence of impurities. Finally, emission
properties are identical at the sub-micron and sub-millimeter scale,
demonstrating the potential of these structures for large scale
applications.

\end{abstract}

\vspace{0.5cm}



Nanowires (NWs) are filamentary crystals with a diameter in the
sub-micrometer down to nanometer range. Their special morphology,
dimensions and high surface-to-volume ratio are often translated
into advantageous optical and electrical properties. As a
consequence, they have been widely used in
electronics~\cite{Hu99,Cui01,Therlander06,Lieber07,Lu07},
optoelectronics\cite{Hua09,Saxena13}, solar
cells\cite{Atwater10,Wallentin13,Krogstrup13,Mann16} and sensors
\cite{Tian12, Takei10}. If not adequately passivated, the surface
recombination can limit the optical performance of the
NWs\cite{Nelson78}. In addition, surface depletion can also affect
the volume distribution and separation of the carriers in the
NW\cite{Schricker06,Parkinson07,Demichel10,Katzenmeyer10,Calarco05}.
Different passivation methods have been employed in the past,
notably capping of the free surfaces with a higher bandgap shell
around the nanowire\cite{Noborisaka05,Chang12,Mallor15}.
Nevertheless, capping also modifies the nature of the surface.
Several effects have been reported, such as band bending at the
interface leading to the accumulation of the charge at the interface
or piezo electric
strain\cite{Yuan85,Dhaka13,Smith10,Songmuang16,Hocevar12}. In
addition, the AlGaAs alloy typically used for capping GaAs nanowires
is generally inhomogeneous, with directed and random segregation of
Ga and Al forming respectively Al-rich ridges and Ga-rich nanoscale
islands ~\cite{Heiss13,Mancini14}. Simultaneously, III-V NWs can
suffer from twin defects and polytypism\cite{Heiss11,Plissard10},
which adversely affect their electronic and optical
properties\cite{Thelander11, deLaMata12,Arbiol09}. With a judicious
optimization of growth conditions, single NWs with a pure
zinc-blende or wurzite structure can be
obtained~\cite{Shtrikman09,Shtrikman09a,Plochocka13}. Still, the
optical and electronic properties tend to fluctuate considerably
from NW to NW, which precludes the proper control of the response of
an ensemble of nanowires.

Recently, alternative approaches to obtain defect-free nano
structures have been proposed. Particularly promising is the
inversion of polarity from B to A as well as template assisted and
nano-membrane assisted selective epitaxy (TASE and MASE,
respectively). All these techniques provide defect free III-V nano
structures by blocking the formation of twinning
defects\cite{Yuan15,Chun13,Arab15,Chang2014,delaMata15,Tutuncuoglu15}.
An additional advantage of these approaches is the possibility to
engineer the shape, so that membranes\cite{Tutuncuoglu15}, sails
\cite{delaMata15} or sheets\cite{Chun13,Chang2014} can be grown.
Nanoscale membranes show relatively long minority carrier diffusion
length of 180\,nm at 4.2\,K, which is significantly larger than the
diffusion lengths found in nanowires \cite{Arab15,Chang2014}.
Moreover, by introducing passivation and/or doped structures, the
design can be further sophisticated \cite{Arab15,Chang2014}. The
transfer of NW optoelectronic devices to industry requires achieving
highly reproducible and uniform structures through a large surface
area, so that the properties of ensemble and single object are
indistinguishable. For instance, in photoluminescence this implies
indistinguishability in terms of line width and emission energy and
spectral shape. Growing the nano structures using TASE and MASE
turned out to be the most promising direction to achieve large area
highly uniform systems.

In this work we demonstrate, by using optical techniques, that GaAs
nanoscale membranes provide the settings for extremely high quality
nanostructures, both from the structural and functional point of
view. We elucidate how the improvement in functional properties is
homogeneous across the whole wafer. This shows the potential of
these nanostructures for nanotechnology and opens the path towards
large scale nano-manufacturing. Furthermore, we provide very strong
evidence that capping of the membranes, despite increasing the
emission efficiency, unexpectedly leads to the degradation of their
optical properties.

\begin{figure}[h]
\includegraphics[width=6cm]{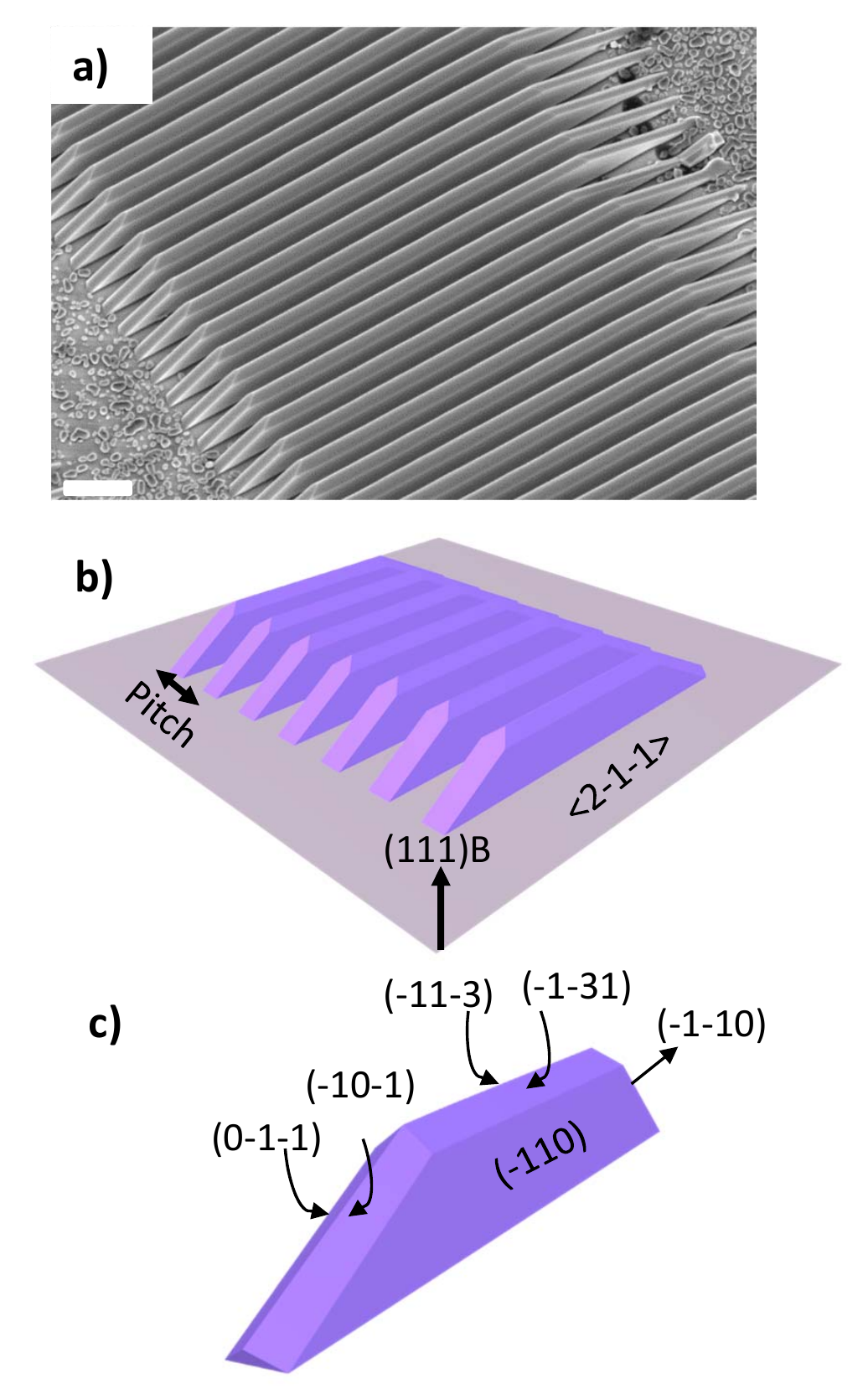}
\caption{(a)  $20^\circ$ tilted SEM image of a GaAs nano membrane
array (b)  3D model of the membrane array signifying the orientation
of the structures (c)  Faceting of a GaAs
membrane.}\label{fig_growth}
\end{figure}

Nanomembranes have been grown using selective area epitaxy (for
growth details see methods and reference\cite{Tutuncuoglu15}). In
Fig.\,\ref{fig_growth}(a) a tilted SEM image of a GaAs nanomembrane
array consisting of $10\mu$m long and $100$\,nm wide nanomembranes
with $500$\,nm pitch, used in the further optical experiments, are
shown. Pitch is defined as the distance between the membranes, as
depicted in Fig.\,\ref{fig_growth}(b). Nanomembranes are oriented in
$\langle2\bar{1}\bar{1}\rangle$ direction which is perpendicular to
$(111)$B and $(\bar{1}\bar{1}0)$ directions and expose the facets
shown in Figure\,\ref{fig_growth}(c). Most of the facets belong to
$\{110\}$ family except high index top facets of $(\bar{1}1\bar{3})$
and $(\bar{1}\bar{3}1)$. Adjusting the membrane orientation to
$\langle2\bar{1}\bar{1}\rangle$ and growth conditions, it is
possible to obtain pure zinc-blende structures with high-aspect
ratio with Molecular Beam Epitaxy (MBE)\cite{Tutuncuoglu15}.
Detailed growth conditions are given in the Method section. The
reported shape is the result of an hour growth with 1\,{\AA}/s
growth rate. If growth is continued long enough, the morphology of
the membrane evolves into a triangular shape. During the growth of
AlGaAs shell the $(\bar{1}\bar{1}0)$ facet transforms to
$(\bar{2}\bar{2}1)$.

\begin{figure}[h]
\includegraphics[width=15cm]{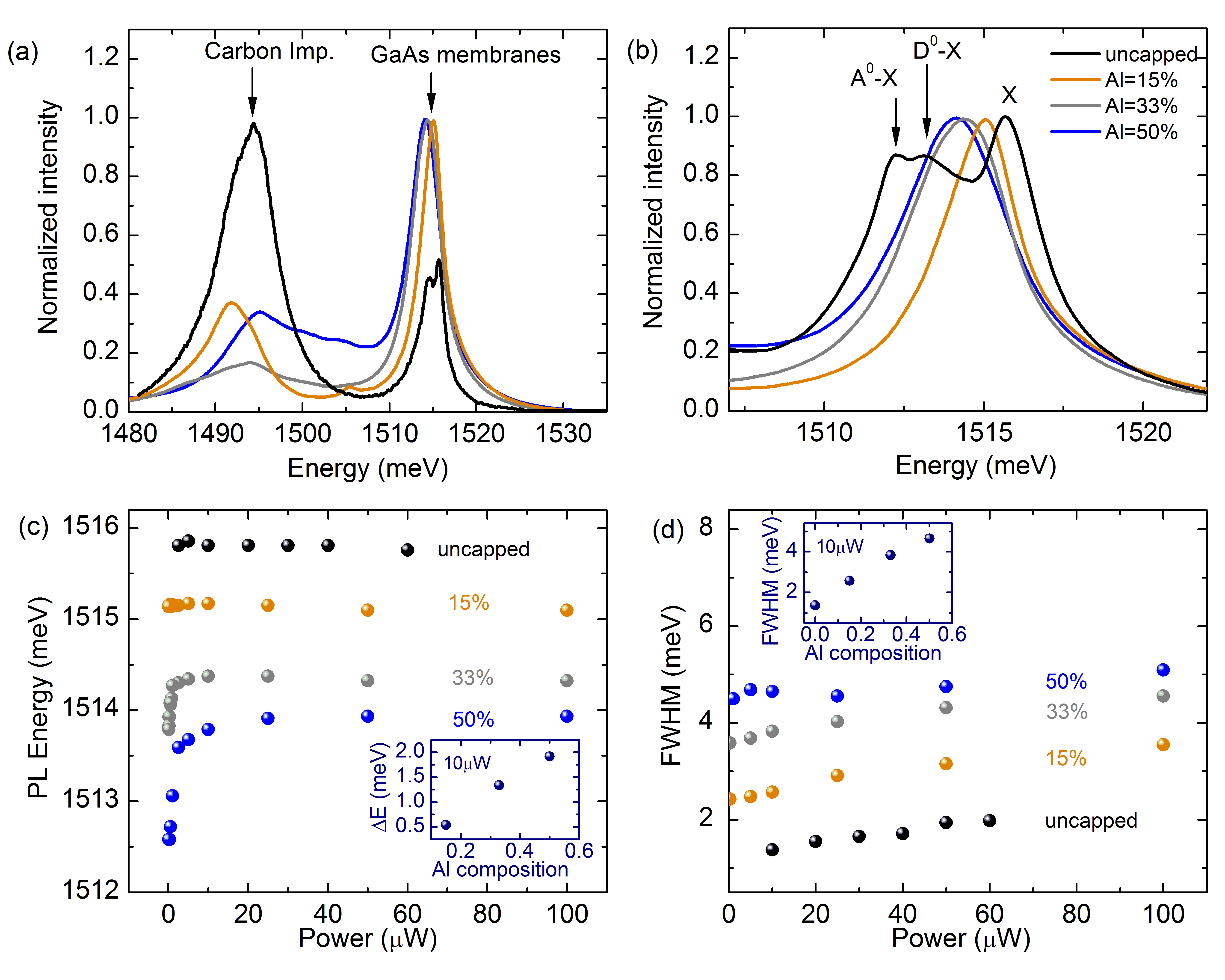}
\caption{ Normalized photoluminescence spectra of the uncapped GaAs
and capped GaAs/AlGaAs nano membranes in a (a) wide (full) and (b)
narrower (emission from the core) energy range. (c) Emission energy
and (d) FWHM of the free exciton emission as a function of the
excitation power. Red-shift of the PL for different Al composition
is plotted as symbols in the inset in panel (c). The inset in panel
(d) shows the evolution of the FWHM of the PL line for different Al
composition in the shell.}\label{fig1}
\end{figure}

Typical normalized $\mu$PL spectra of a single uncapped GaAs and
capped GaAs/Al$_{x}$Ga$_{1-x}$As nano membranes are presented in
Fig.\,\ref{fig1}(a). For the capped membranes the data have been
taken for three different compositions of the shell
($x=15,33,50\%$). Overall, the emission spectra are composed of two
bands, around $1515$ and $1490$ meV. The higher energy band
corresponds to the band-edge luminescence of the GaAs membranes,
while the lower energy emission can be attributed to the donor
acceptor transitions due to carbon impurities normally present in
commercial GaAs substrates\cite{Heiß08,Woolf92}, which was further
observed in detailed cathodoluminescence studies. Our spectra are
comparable to previously reported optical emission in nano membranes
with $33\%$ Al composition in the shell\cite{Tutuncuoglu15}. The
peak related to the carbon impurities can be used as a reference for
the luminescence intensity. After capping, the emission from the
GaAs membrane increases dominating the carbon related PL.  The
dramatic increase of the emission from the membrane is a direct
consequence of the surface passivation that reduces the non
radiative surface recombination.

The detailed nature of the emission is very different for capped and
uncapped samples. For uncapped membranes the spectrum is composed of
three lines (see Fig\ref{fig1}(b)). The peak at the highest energy
of $\sim1515.5$ meV corresponds to the free exciton emission, while
the two peaks at lower energies are related to neutral donor bound
exciton emission ($D^{0}-X$) and acceptor bound exciton emission
($A^{0}-X$) with emission energies which are typical for bulk GaAs
~\cite{Heim74,Kunzel80}. This result rules out any possible quantum
confinement in the nano membranes. This is not unexpected since the
exciton Bohr radius of $\simeq 14$\,nm in GaAs is much smaller than
the size of the membrane\cite{Zhang09}. In contrast, the typical
emission spectra for the GaAs nano membranes capped with
Al$_{x}$Ga$_{1-x}$As layer (Fig.\,\ref{fig1}(b)) are composed of a
single line, which we attribute to the neutral exciton
recombination. Emission lines from $D^{0}-X$ and $A^{0}-X$ are
completely absent. The neutral exciton emission energy red shifts
and broadens with increasing Al shell content. To quantify this
effect we have measured the power dependence of the energy and full
width at the half maximum (FWHM) of the neutral exciton emission. In
Fig~\ref{fig1}(c) the emission energy is plotted as a function of
excitation power. For membranes with high aluminium shell content
($x \geq 0.3$) a blue shift is observed with increasing excitation
power which quickly saturates for powers above a few $\mu$W. For
powers of $10\mu$W and above the emission energy is independent of
the excitation power. There is a clear and systematic decrease in
the emission energy (red-shift) with increasing Al content. This is
illustrated in the inset in the Fig.\,\ref{fig1}(c), where the
energy difference between uncapped and capped emission $\Delta E$ is
plotted as a function of the shell aluminium composition $x$ for the
same excitation power. In Fig.\,\ref{fig1}(d) the FWHM of the
emission is plotted as a function of the excitation power. The line
widths increase slightly with increasing power, but this is
negligible compared to the increase in the FWHM with increasing Al
content of the shell. In the inset of Fig.\,\ref{fig1}(d) we plot
the FWHM versus the shell Al content for an excitation power of
$10\mu$W. The linewidth is multiplied by roughly a factor of 5
between the uncapped membrane and the membrane with a 50\% Al
content cap layer. Thus, while capping the membranes reduces
non-radiative surface recombination, leading to enhanced neutral
exciton emission, it also detrimentally affects the optical
properties of the GaAs core, leading to a significantly broadened
emission.

We turn now to the effect of the red-shift of the excitonic emission
upon capping the membranes with AlGaAs. In fact, a similar effect
has been observed previously for a simple AlGaAs/GaAs
interface~\cite{Ploog83,Yuan85}, InP nanowires, \cite{Weert06} and
for GaAs nanowires capped with AlGaAs shell
~\cite{Dhaka13,Songmuang16,Hocevar12}. For simple AlGaAs/GaAs the
band bending at the interface forms a pocket for the electrons or
holes ~\cite{Ploog83,Yuan85}. Such confined carriers at the
interface will recombine with the free carriers (of the opposite
species) in the valence or conduction band at a sufficient distance
from the interface that flat-band conditions have been
re-established. As the charges are spatially separated, emission has
a spatially indirect character and is red shifted in comparison to
the simple excitonic emission observed in uncapped GaAs. Moreover,
the band bending can be screened by photo created carriers
decreasing the overall effect with the increase of the excitation
power. For InP nanowires a similar picture has been proposed, where
the band bending was induced by a pinning of the Fermi
level\cite{Weert06}. Finally, for GaAs nanowires capped with AlGaAs
shell, the mechanism of the band bending can be enriched by strain,
related to the shell thickness~\cite{Songmuang16,Hocevar12}.
However, the strain plays a significant role only for rather thick
shells. In the case of the nano membranes the core is much thicker
than the shell. Additional confirmation of the negligible role of
the strain in our structures is given by the Raman spectroscopy. If
the shift we observe originated from strain, it would imply a
significantly lower Al composition than the nominal composition
\cite{Signorell013}. Our Raman measurements, (see SI), confirm that
the Al composition corresponds very well to the nominal composition
in the nano membranes and the lack of strain in the membrane core.

\begin{figure}[h]
\includegraphics[width=15cm]{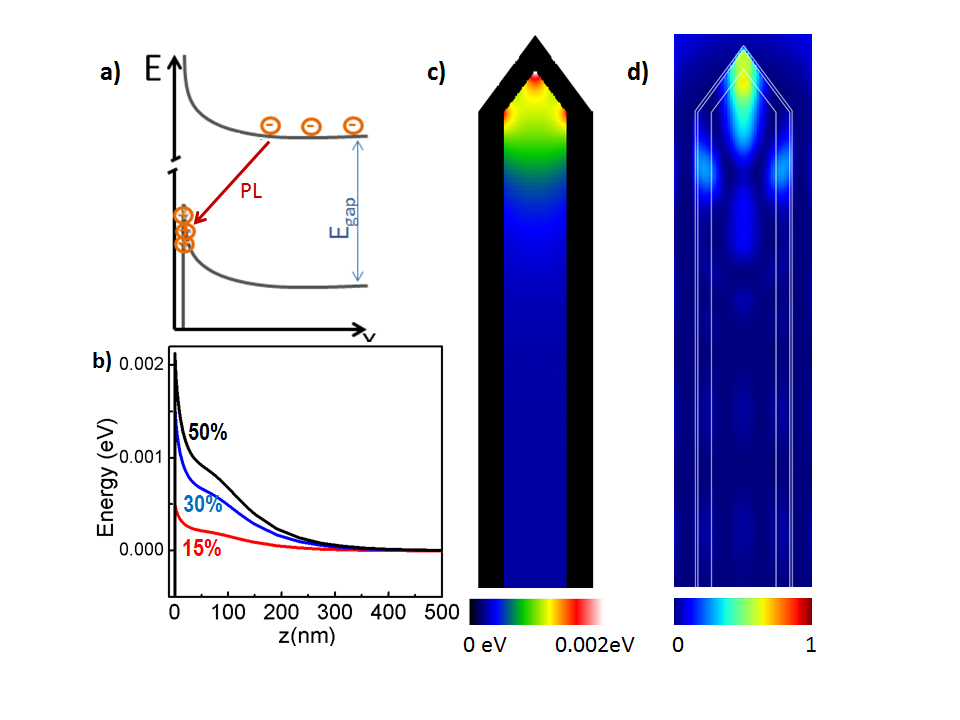}
\caption{(a) The position of the valence and conduction band edges
as a function of the distance from the membrane surface. b) Band
bending at the tip of the membrane as a function of the distance to
the surface and for the three Al contents. (c) 2D valence and the
conduction band maps for an Al concentration of $50\%$ (d)
Cross-sectional map of the electric field energy density for a
nanomembrane under the presence of a monochromatic wave coming from
the top and with parallel polarization }\label{figSimulation}
\end{figure}

We attribute the observed red shift of the emission to the indirect
nature of the exciton recombination at the capping interface. Due to
residual doping in the AlGaAs shell, band bending occurs at the
AlGaAs/GaAs interface. To this end, we illustrate in Fig 3(a) the
position of the valence and conduction band edges as a function of
the distance from the membrane surface. Our hypothesis is that the
AlGaAs shell contains some oxygen impurities, associated with the
addition of aluminum. Secondary ion mass spectroscopy measurements
on AlGaAs layers indeed indicate a slight O-contamination associated
with Aluminum (see SI). This contamination is still better than the
purity specifications of MBE-grade Aluminum, 6N5', which implies
that nanostructures are much more sensitive than bulk structures to
impurities. Thus, the optical response of high quality nano
structures provides a sensitive means to detects extremely low
levels of impurities. The red-shift of the luminescence at high
excitation powers is larger for higher Al content (see Figure
\ref{figSimulation}(a)). This shows, that the band bending increases
with the Al content in the shell as the exciton recombination
becomes more indirect.

Our observations are further supported by the simulation of the band
bending at the AlGaAs/GaAs interface by solving Poisson and
Schr\"odinger equations self-consistently with the software
nextnano3. In the model we have included the presence of p-type
interface states between GaAs and AlGaAs shell, which increases with
increasing Al content. Our experimental data fits well with $2
\times 10^9$, $6 \times 10^9$ and $8 \times 10^9$\,cm$^{-2}$
interface dopants for an Al concentration of 15\%, 30\% and 50\%
respectively. Fig.\,\ref{figSimulation}(b) shows the resulting band
bending at the tip of the membrane as a function of the distance to
the surface and for the three Al contents. Here is evident the
presence of a triangular potential in the valence band at the
interface GaAs/AlGaAs where holes can be trapped. We can also
observe an increase of the height of the potential with Al content,
which results in a red-shift of the indirect transition.

It is worth noting, that the red shift observed in our samples is of
comparable magnitude with that observed by Songmuang at
al~\cite{Songmuang16} but much smaller than that reported by Dhaka
et al~~\cite{Dhaka13}. This discrepancy can be partly ascribed to
the MetalOrganic Vapor Phase Epitaxy (MOVPE) employed by Dhaka et al
~\cite{Dhaka13} to grow their nanowires.  MOVPE involves the use of
metalorganic species as group III precursors, which might introduce
an unintentionally high concentration of impurities.

The small blue shift observed at low powers, which saturates around
$10 \mu$W has been also observed for GaAs nanowires capped with
AlGaAs shell \cite{Songmuang16,Hocevar12} and it was associated with
the presence of some negatively charged traps at the interface,
which are filled by photo created carriers in the AlGaAs shell,
which migrates towards interface. Once filled, they can no longer
modify the band bending at the interface, which explains the
saturation of the blue-shift of the emission energy above $10 \mu$W,
indicating that the band bending is the dominant effect in our
nanomembranes.

The special geometry of the nanoscale membranes requires some
further modeling. First, the non-flat geometry of the interface
should result into a spatially dependent band bending. In addition,
the vertical nature of the membranes can additionally lead to
non-homogeneous light absorption~\cite {Heiss13}.
Fig.\,\ref{figSimulation}(c) shows the 2D valence and the conduction
band maps for an Al concentration of 50\%. We can observe a
band-bending at the interface which is significantly larger at the
top corner of the nanomembrane. We have simulated the
electromagnetic field distribution using the package Meep, a freely
available software implementing the Finite Difference in Time Domain
Method ~\cite{Oksooi10} taking into consideration the exact geometry
of the core/shell nanomembrane with a shell of 30\% of Al. The
dielectric constant is taken from Ref.\,[\cite{Adachi1985}].
Fig.\,\ref{figSimulation}(d) shows the cross-sectional map of the
computed electric field energy density for a nanomembrane under the
presence of a monochromatic wave coming from the top and with
parallel polarization. It is clearly seen that the field energy is
not distributed evenly across the cross-section but is rather
confined at the top edge of the nanomembrane. This means that our
$\mu$ photoluminescence experiments mainly probe the exciton
properties at the tip, where the band bending is more pronounced.
The results of this simulation explains also the broadening of the
emission with the increasing Al content. Although emission is probed
locally, the probed region can contain non homogenies band bending
leading to the broadening of the emission peak. This is in perfect
agreement with the observation that the effect is the strongest for
the highest Al composition.

Finally, we come to perhaps the most striking and novel property of
these nano membranes, namely their reproducibility and large scale
uniformity.  While epitaxial MBE provides highly uniform growth,
this is not the case for the self organized growth of quantum dots
or NWs, where nucleation events in growth follow poissonian
statistics that lead to a distribution in the properties. As an
example, in NWs this leads to a twinning or stacking fault density
that varies from NW to NW (complete defect-free structures are
rare). As a result, the optical properties vary from NW to NW and
macro-photoluminescence measurements of the ensemble normally do not
match micro photoluminescence of a single NW. We have recently shown
that MBE growth using selective area epitaxy can produce arrays of
defect free nano membranes  \cite{Tutuncuoglu15}. However, optical
investigations were limited to PL of a single nano membrane so that
the uniform optical properties of an ensemble has never been
demonstrated.

\begin{figure}[h]
\includegraphics[width=12cm]{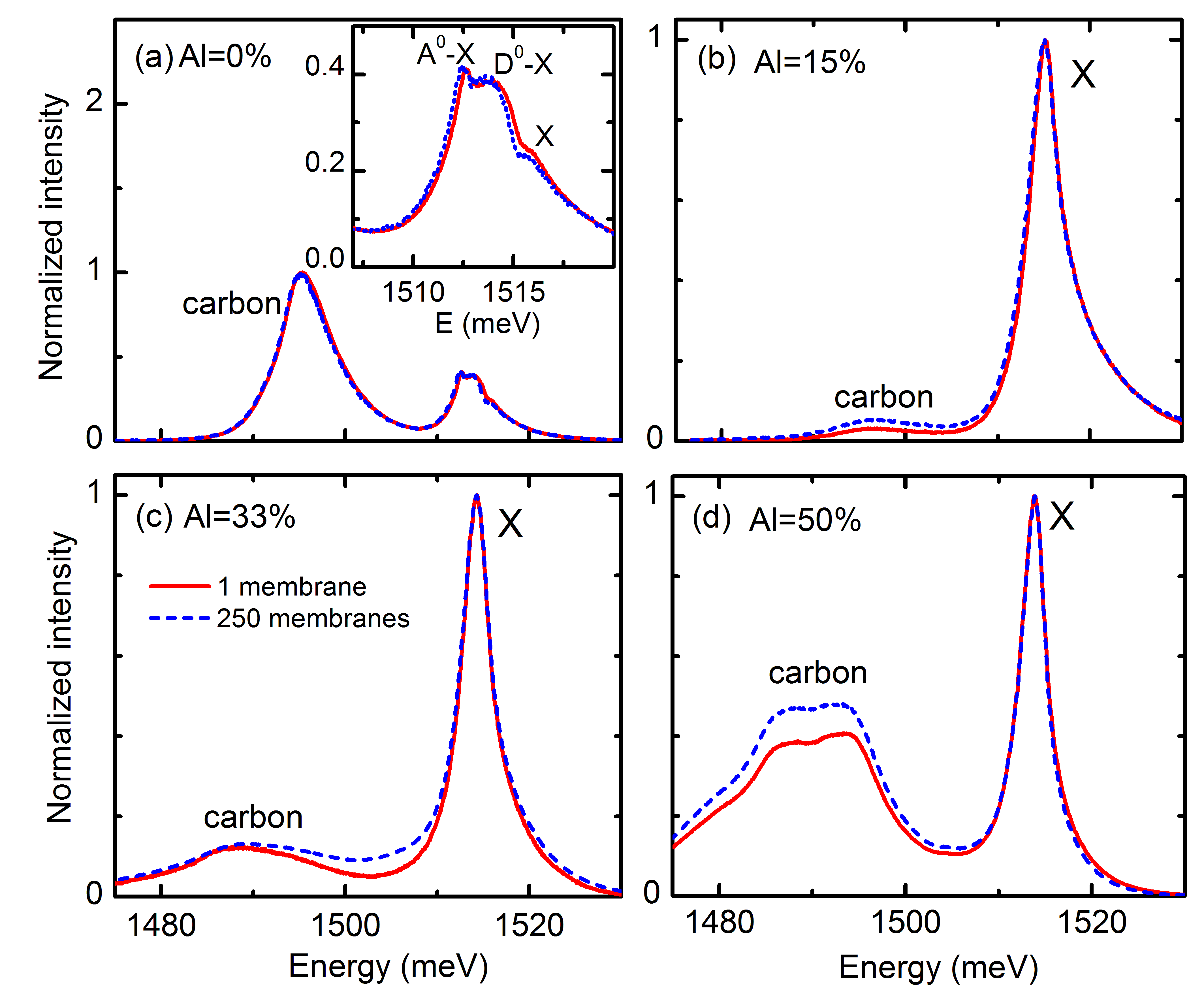}
\caption{(a)-(d) Normalized photoluminescence spectra of the
uncapped GaAs and capped GaAs/AlGaAs nano membranes measured at low
temperature.  The solid red and dashed blue line correspond to the
excitation/emission from 1 and 250 membranes respectively. The inset
shows the zoom of the GaAs core emission of the uncapped
membranes.}\label{macro}
\end{figure}

To demonstrate large scale uniformity, we compare the emission
spectra of a single membrane with the ensemble emission of around
$250$ membranes measured using macro PL, achieved here by
defocussing the laser spot. Representative PL spectra are shown in
Fig.\,\ref{macro} for the capped and uncapped membranes. Defocussing
increases the contribution of the substrate which is reflected in
the slightly increased amplitude of the carbon related emission
which can be seen in Figure\,\ref{macro}. The substrate PL is
dominated by the carbon related emission and free exciton emission
is not observed from the substrate.  We have mapped the luminescence
properties of the membranes by cathodoluminescence in a previous
work \cite{Tutuncuoglu15}. These measurements confirm that the
carbon-related peak originates solely from the substrate.

Surprisingly, the PL originating from single membrane is almost identical to the ensemble emission. The carbon impurity emission
is slightly enhanced in the ensemble emission of the capped samples ($\simeq 20$\% for the 50\% Al membrane). This is probably
due to the inhomogeneous distribution of the carbon impurities across the substrate. In contrast, the neutral exciton emission is
strictly identical in both the energy of the emission and the line width for all samples. In the uncapped sample, the neutral and
bound exciton emission is also almost indistinguishable (see inset Figure\,\ref{macro}(a)). The identical emission from a single
and an ensemble of membranes unequivocally demonstrates the very high quality of the crystal structure and extremely high
reproducibility of the nano membranes, which has never been observed for the classical radial nanowires. Moreover, data measured
at different places on the same membrane, and on different membranes, vary very little in intensity, energy position or line
width, suggesting an excellent crystal quality of the uncapped membranes.

In conclusion, we have demonstrated luminescence properties of GaAs
membranes which are on a par with the best two-dimensional layers
obtained with MBE.  Upon capping of the membranes with an AlGaAs
layer the PL emission is strongly enhanced, but also unexpectedly
accompanied by a degradation of the optical properties with a
significant broadening of the exciton emission. Capping also leads
to a red shift of the emission which has been attributed to the
residual carbon doping of Al-containing layers which leads to band
bending at the AlGaAs/GaAs interface. The quality of the membrane
growth process is further supported by ensemble measurements, which
are almost indistinguishable from the single membrane results.
Additionally, our results show an extreme high sensitivity of the
optical response of the nano membranes on impurities concentration
that goes beyond what is possible in terms of state of the art high
purity MBE.

ASSOCIATED CONTENT

Additional characterization of membranes, inelastic light scattering (Raman), secondary ion mass spectroscopy (SIMS) profile
(PDF).

ACKNOWLEDGEMENTS

This work was partially supported by ANR JCJC project milliPICS, the
R\'egion Midi-Pyr\'en\'ees under contract MESR 13053031, BLAPHENE
project under IDEX program Emergence

\section{Methods}

GaAs nanomembranes used in that study are grown with a DCA solid
source Molecular Beam Epitaxy (MBE) system. Substrates are PECVD
deposited SiO$_2$ masked $(111)$B GaAs. The oxide thickness is
30\,nm. The growth mask is patterned with a combination of e-beam
lithography and fluorine based dry etching as reported earlier
\cite{Tutuncuoglu15}. The growth temperature is 635 $^\circ$C, the
growth rate is 1\,{\AA}/s and the V/III ratio is 10 for the GaAs
core. The length of nanomembranes and the distance between them are
defined by patterning the SiO$_{2}$ mask. We focused our
characterization on structures having structures with 100\,nm width,
500\,nm pitch and $10\mu$m length are studied. In the case of capped
GaAs nanomembranes, the structures are capped with a shell of
Al$_x$Ga$_{1-x}$As. ( $x = 0.15, 0.3$ and $0.5$) The substrate
temperature is reduced to 460 $^\circ$C and As flux is increased to
$1 \times 10^{-5}$\,torr for shell growth. Nominal thickness of
AlGaAs shell is always 50\,nm and it is protected with 10\,nm of
GaAs against oxidization. Aluminum ratios and nominal AlGaAs layer
thicknesses are deduced from RHEED calibrations performed on $(100)$
GaAs substrates.

For the measurements the samples were placed in a helium flow
cryostat with optical access. The cryostat was mounted on the
motorized $x-y$ translation stages, which allows high resolution
spatial mapping. A microscope objective 50$\times$  with a numerical
aperture NA\,=\,0.55 was used to focus the excitation beam and
collect the PL from the nano membranes. The laser spot could be
focussed down to a diameter of $\simeq 0.5\mu$m (diffraction limit),
which enabled us to optically address single (or a maximum of two in
a worst case scenario). To investigate many membranes the laser spot
was defocussed. The steady-state $\mu$PL signal was excited with a
532\,nm laser and the spectra were recorded using a spectrometer
equipped with a liquid nitrogen cooled CCD camera. All the
measurements presented here have been performed at $T=4.2$\,K.

\bibliography{Nws}

\providecommand{\latin}[1]{#1}
\makeatletter
\providecommand{\doi}
  {\begingroup\let\do\@makeother\dospecials
  \catcode`\{=1 \catcode`\}=2\doi@aux}
\providecommand{\doi@aux}[1]{\endgroup\texttt{#1}}
\makeatother
\providecommand*\mcitethebibliography{\thebibliography}
\csname @ifundefined\endcsname{endmcitethebibliography}
  {\let\endmcitethebibliography\endthebibliography}{}
\begin{mcitethebibliography}{54}
\providecommand*\natexlab[1]{#1}
\providecommand*\mciteSetBstSublistMode[1]{}
\providecommand*\mciteSetBstMaxWidthForm[2]{}
\providecommand*\mciteBstWouldAddEndPuncttrue
  {\def\EndOfBibitem{\unskip.}}
\providecommand*\mciteBstWouldAddEndPunctfalse
  {\let\EndOfBibitem\relax}
\providecommand*\mciteSetBstMidEndSepPunct[3]{}
\providecommand*\mciteSetBstSublistLabelBeginEnd[3]{}
\providecommand*\EndOfBibitem{}
\mciteSetBstSublistMode{f}
\mciteSetBstMaxWidthForm{subitem}{(\alph{mcitesubitemcount})}
\mciteSetBstSublistLabelBeginEnd
  {\mcitemaxwidthsubitemform\space}
  {\relax}
  {\relax}

\bibitem[Hu \latin{et~al.}(1999)Hu, Odom, and Lieber]{Hu99}
Hu,~J.; Odom,~T.~W.; Lieber,~C.~M. \emph{Accounts of Chemical Research}
  \textbf{1999}, \emph{32}, 435--445\relax
\mciteBstWouldAddEndPuncttrue
\mciteSetBstMidEndSepPunct{\mcitedefaultmidpunct}
{\mcitedefaultendpunct}{\mcitedefaultseppunct}\relax
\EndOfBibitem
\bibitem[Cui and Lieber(2001)Cui, and Lieber]{Cui01}
Cui,~Y.; Lieber,~C.~M. \emph{Science} \textbf{2001}, \emph{291}, 851\relax
\mciteBstWouldAddEndPuncttrue
\mciteSetBstMidEndSepPunct{\mcitedefaultmidpunct}
{\mcitedefaultendpunct}{\mcitedefaultseppunct}\relax
\EndOfBibitem
\bibitem[Thelander(2006)]{Therlander06}
Thelander,~C. \emph{Mater. Today} \textbf{2006}, \emph{9}, 28\relax
\mciteBstWouldAddEndPuncttrue
\mciteSetBstMidEndSepPunct{\mcitedefaultmidpunct}
{\mcitedefaultendpunct}{\mcitedefaultseppunct}\relax
\EndOfBibitem
\bibitem[Lieber and Wang(2007)Lieber, and Wang]{Lieber07}
Lieber,~C.~M.; Wang,~Z.~L. \emph{MRS Bulletin} \textbf{2007}, \emph{32},
  99--108\relax
\mciteBstWouldAddEndPuncttrue
\mciteSetBstMidEndSepPunct{\mcitedefaultmidpunct}
{\mcitedefaultendpunct}{\mcitedefaultseppunct}\relax
\EndOfBibitem
\bibitem[Lu and Lieber(2007)Lu, and Lieber]{Lu07}
Lu,~W.; Lieber,~C.~M. \emph{Nature Mater.} \textbf{2007}, \emph{6}, 841\relax
\mciteBstWouldAddEndPuncttrue
\mciteSetBstMidEndSepPunct{\mcitedefaultmidpunct}
{\mcitedefaultendpunct}{\mcitedefaultseppunct}\relax
\EndOfBibitem
\bibitem[Hua \latin{et~al.}(2009)Hua, Motohisa, Kobayashi, Hara, and
  Fukui]{Hua09}
Hua,~B.; Motohisa,~J.; Kobayashi,~Y.; Hara,~S.; Fukui,~T. \emph{Nano Letters}
  \textbf{2009}, \emph{9}, 112--116, PMID: 19072060\relax
\mciteBstWouldAddEndPuncttrue
\mciteSetBstMidEndSepPunct{\mcitedefaultmidpunct}
{\mcitedefaultendpunct}{\mcitedefaultseppunct}\relax
\EndOfBibitem
\bibitem[Saxena \latin{et~al.}({2013})Saxena, Mokkapati, Parkinson, Jiang, Gao,
  Tan, and Jagadish]{Saxena13}
Saxena,~D.; Mokkapati,~S.; Parkinson,~P.; Jiang,~N.; Gao,~Q.; Tan,~H.~H.;
  Jagadish,~C. \emph{{Nature Photonics}} \textbf{{2013}}, \emph{{7}},
  {963--968}\relax
\mciteBstWouldAddEndPuncttrue
\mciteSetBstMidEndSepPunct{\mcitedefaultmidpunct}
{\mcitedefaultendpunct}{\mcitedefaultseppunct}\relax
\EndOfBibitem
\bibitem[Atwater and Polman({2010})Atwater, and Polman]{Atwater10}
Atwater,~H.~A.; Polman,~A. \emph{{Nature Materials}} \textbf{{2010}},
  \emph{{9}}, {205--213}\relax
\mciteBstWouldAddEndPuncttrue
\mciteSetBstMidEndSepPunct{\mcitedefaultmidpunct}
{\mcitedefaultendpunct}{\mcitedefaultseppunct}\relax
\EndOfBibitem
\bibitem[Wallentin \latin{et~al.}(2013)Wallentin, Anttu, Asoli, Huffman, {\r
  A}berg, Magnusson, Siefer, Fuss-Kailuweit, Dimroth, Witzigmann, Xu,
  Samuelson, Deppert, and Borgstr{\"o}m]{Wallentin13}
Wallentin,~J.; Anttu,~N.; Asoli,~D.; Huffman,~M.; {\r A}berg,~I.;
  Magnusson,~M.~H.; Siefer,~G.; Fuss-Kailuweit,~P.; Dimroth,~F.;
  Witzigmann,~B.; Xu,~H.~Q.; Samuelson,~L.; Deppert,~K.; Borgstr{\"o}m,~M.~T.
  \emph{Science} \textbf{2013}, \emph{339}, 1057--1060\relax
\mciteBstWouldAddEndPuncttrue
\mciteSetBstMidEndSepPunct{\mcitedefaultmidpunct}
{\mcitedefaultendpunct}{\mcitedefaultseppunct}\relax
\EndOfBibitem
\bibitem[Krogstrup \latin{et~al.}(2013)Krogstrup, J{\o}rgensen, Heiss,
  Demichel, Holm, Aagesen, Nygard, and i~Morral]{Krogstrup13}
Krogstrup,~P.; J{\o}rgensen,~H.~I.; Heiss,~M.; Demichel,~O.; Holm,~J.~V.;
  Aagesen,~M.; Nygard,~J.; i~Morral,~A.~F. \emph{Nature Photonics}
  \textbf{2013}, \emph{7}, 306--310\relax
\mciteBstWouldAddEndPuncttrue
\mciteSetBstMidEndSepPunct{\mcitedefaultmidpunct}
{\mcitedefaultendpunct}{\mcitedefaultseppunct}\relax
\EndOfBibitem
\bibitem[Mann \latin{et~al.}(2016)Mann, Oener, Cavalli, Haverkort, Bakkers, and
  Garnett]{Mann16}
Mann,~S.~A.; Oener,~S.~Z.; Cavalli,~A.; Haverkort,~J. E.~M.; Bakkers,~E. P.
  A.~M.; Garnett,~E.~C. \emph{Nat Nano} \textbf{2016}, \emph{advance online
  publication}\relax
\mciteBstWouldAddEndPuncttrue
\mciteSetBstMidEndSepPunct{\mcitedefaultmidpunct}
{\mcitedefaultendpunct}{\mcitedefaultseppunct}\relax
\EndOfBibitem
\bibitem[Tian \latin{et~al.}(2012)Tian, Liu, Dvir, Jin, Tsui, Qing, Suo,
  Langer, Kohane, and Lieber]{Tian12}
Tian,~B.; Liu,~J.; Dvir,~T.; Jin,~L.; Tsui,~J.~H.; Qing,~Q.; Suo,~Z.;
  Langer,~R.; Kohane,~D.~S.; Lieber,~C.~M. \emph{Nat Mater} \textbf{2012},
  \emph{11}, 986--994\relax
\mciteBstWouldAddEndPuncttrue
\mciteSetBstMidEndSepPunct{\mcitedefaultmidpunct}
{\mcitedefaultendpunct}{\mcitedefaultseppunct}\relax
\EndOfBibitem
\bibitem[Takei \latin{et~al.}(2010)Takei, Takahashi, Ho, Ko, Gillies, Leu,
  Fearing, and Javey]{Takei10}
Takei,~K.; Takahashi,~T.; Ho,~J.~C.; Ko,~H.; Gillies,~A.~G.; Leu,~P.~W.;
  Fearing,~R.~S.; Javey,~A. \emph{Nat Mater} \textbf{2010}, \emph{9},
  821--826\relax
\mciteBstWouldAddEndPuncttrue
\mciteSetBstMidEndSepPunct{\mcitedefaultmidpunct}
{\mcitedefaultendpunct}{\mcitedefaultseppunct}\relax
\EndOfBibitem
\bibitem[Nelson and Sobers(1978)Nelson, and Sobers]{Nelson78}
Nelson,~R.~J.; Sobers,~R.~G. \emph{Applied Physics Letters} \textbf{1978},
  \emph{32}, 761--763\relax
\mciteBstWouldAddEndPuncttrue
\mciteSetBstMidEndSepPunct{\mcitedefaultmidpunct}
{\mcitedefaultendpunct}{\mcitedefaultseppunct}\relax
\EndOfBibitem
\bibitem[Schricker \latin{et~al.}(2006)Schricker, III, Wiacek, and
  Korgel]{Schricker06}
Schricker,~A.~D.; III,~F. M.~D.; Wiacek,~R.~J.; Korgel,~B.~A.
  \emph{Nanotechnology} \textbf{2006}, \emph{17}, 2681\relax
\mciteBstWouldAddEndPuncttrue
\mciteSetBstMidEndSepPunct{\mcitedefaultmidpunct}
{\mcitedefaultendpunct}{\mcitedefaultseppunct}\relax
\EndOfBibitem
\bibitem[Parkinson \latin{et~al.}(2007)Parkinson, Lloyd-Hughes, Gao, Tan,
  Jagadish, Johnston, and Herz]{Parkinson07}
Parkinson,~P.; Lloyd-Hughes,~J.; Gao,~Q.; Tan,~H.~H.; Jagadish,~C.;
  Johnston,~M.~B.; Herz,~L.~M. \emph{Nano Letters} \textbf{2007}, \emph{7},
  2162--2165\relax
\mciteBstWouldAddEndPuncttrue
\mciteSetBstMidEndSepPunct{\mcitedefaultmidpunct}
{\mcitedefaultendpunct}{\mcitedefaultseppunct}\relax
\EndOfBibitem
\bibitem[Demichel \latin{et~al.}(2010)Demichel, Heiss, Bleuse, Mariette, and
  Fontcuberta~i Morral]{Demichel10}
Demichel,~O.; Heiss,~M.; Bleuse,~J.; Mariette,~H.; Fontcuberta~i Morral,~A.
  \emph{Applied Physics Letters} \textbf{2010}, \emph{97}\relax
\mciteBstWouldAddEndPuncttrue
\mciteSetBstMidEndSepPunct{\mcitedefaultmidpunct}
{\mcitedefaultendpunct}{\mcitedefaultseppunct}\relax
\EndOfBibitem
\bibitem[Katzenmeyer \latin{et~al.}(2010)Katzenmeyer, Léonard, Talin, Wong, and
  Huffaker]{Katzenmeyer10}
Katzenmeyer,~A.~M.; Léonard,~F.; Talin,~A.~A.; Wong,~P.~S.; Huffaker,~D.~L.
  \emph{Nano Letters} \textbf{2010}, \emph{10}, 4935--4938, PMID:
  21053980\relax
\mciteBstWouldAddEndPuncttrue
\mciteSetBstMidEndSepPunct{\mcitedefaultmidpunct}
{\mcitedefaultendpunct}{\mcitedefaultseppunct}\relax
\EndOfBibitem
\bibitem[Calarco \latin{et~al.}(2005)Calarco, Marso, Richter, Aykanat, Meijers,
  v.d. Hart, Stoica, and Lüth]{Calarco05}
Calarco,~R.; Marso,~M.; Richter,~T.; Aykanat,~A.~I.; Meijers,~R.; v.d.
  Hart,~A.; Stoica,~T.; Lüth,~H. \emph{Nano Lett.} \textbf{2005}, \emph{5},
  981--984\relax
\mciteBstWouldAddEndPuncttrue
\mciteSetBstMidEndSepPunct{\mcitedefaultmidpunct}
{\mcitedefaultendpunct}{\mcitedefaultseppunct}\relax
\EndOfBibitem
\bibitem[Noborisaka \latin{et~al.}(2005)Noborisaka, Motohisa, Hara, and
  Fukui]{Noborisaka05}
Noborisaka,~J.; Motohisa,~J.; Hara,~S.; Fukui,~T. \emph{Applied Physics
  Letters} \textbf{2005}, \emph{87}\relax
\mciteBstWouldAddEndPuncttrue
\mciteSetBstMidEndSepPunct{\mcitedefaultmidpunct}
{\mcitedefaultendpunct}{\mcitedefaultseppunct}\relax
\EndOfBibitem
\bibitem[Chang \latin{et~al.}(2012)Chang, Chi, Yao, Huang, Chen, Theiss,
  Bushmaker, LaLumondiere, Yeh, Povinelli, Zhou, Dapkus, and Cronin]{Chang12}
Chang,~C.~C.; Chi,~C.-Y.; Yao,~M.; Huang,~N.; Chen,~C.~C.; Theiss,~J.;
  Bushmaker,~A.~W.; LaLumondiere,~S.; Yeh,~T.~W.; Povinelli,~M.~L.; Zhou,~C.;
  Dapkus,~P.~D.; Cronin,~S.~B. \emph{Nano Letters} \textbf{2012}, \emph{12},
  4484--4489, PMID: 22889241\relax
\mciteBstWouldAddEndPuncttrue
\mciteSetBstMidEndSepPunct{\mcitedefaultmidpunct}
{\mcitedefaultendpunct}{\mcitedefaultseppunct}\relax
\EndOfBibitem
\bibitem[Mallorqu{\'i} \latin{et~al.}(2015)Mallorqu{\'i},
  Alarc{\'o}n-Llad{\'o}, Mundet, Kiani, Demaurex, De~Wolf, Menzel, Zacharias,
  and Fontcuberta~i Morral]{Mallor15}
Mallorqu{\'i},~A.~D.; Alarc{\'o}n-Llad{\'o},~E.; Mundet,~I.~C.; Kiani,~A.;
  Demaurex,~B.; De~Wolf,~S.; Menzel,~A.; Zacharias,~M.; Fontcuberta~i
  Morral,~A. \emph{Nano Research} \textbf{2015}, \emph{8}, 673--681\relax
\mciteBstWouldAddEndPuncttrue
\mciteSetBstMidEndSepPunct{\mcitedefaultmidpunct}
{\mcitedefaultendpunct}{\mcitedefaultseppunct}\relax
\EndOfBibitem
\bibitem[Yuan \latin{et~al.}(1985)Yuan, Pudensi, Vawter, and Merz]{Yuan85}
Yuan,~Y.~R.; Pudensi,~M. A.~A.; Vawter,~G.~A.; Merz,~J.~L. \emph{Journal of
  Applied Physics} \textbf{1985}, \emph{58}, 397--403\relax
\mciteBstWouldAddEndPuncttrue
\mciteSetBstMidEndSepPunct{\mcitedefaultmidpunct}
{\mcitedefaultendpunct}{\mcitedefaultseppunct}\relax
\EndOfBibitem
\bibitem[Dhaka \latin{et~al.}(2013)Dhaka, Oksanen, Jiang, Haggren, Nyk\"anen,
  Sanatinia, Kakko, Huhtio, Mattila, Ruokolainen, Anand, Kauppinen, and
  Lipsanen]{Dhaka13}
Dhaka,~V.; Oksanen,~J.; Jiang,~H.; Haggren,~T.; Nyk\"anen,~A.; Sanatinia,~R.;
  Kakko,~J.-P.; Huhtio,~T.; Mattila,~M.; Ruokolainen,~J.; Anand,~S.;
  Kauppinen,~E.; Lipsanen,~H. \emph{Nano Letters} \textbf{2013}, \emph{13},
  3581--3588\relax
\mciteBstWouldAddEndPuncttrue
\mciteSetBstMidEndSepPunct{\mcitedefaultmidpunct}
{\mcitedefaultendpunct}{\mcitedefaultseppunct}\relax
\EndOfBibitem
\bibitem[Smith \latin{et~al.}(2010)Smith, Jackson, Yarrison-Rice, and
  Jagadish]{Smith10}
Smith,~L.~M.; Jackson,~H.~E.; Yarrison-Rice,~J.~M.; Jagadish,~C.
  \emph{Semiconductor Science and Technology} \textbf{2010}, \emph{25},
  024010\relax
\mciteBstWouldAddEndPuncttrue
\mciteSetBstMidEndSepPunct{\mcitedefaultmidpunct}
{\mcitedefaultendpunct}{\mcitedefaultseppunct}\relax
\EndOfBibitem
\bibitem[Songmuang \latin{et~al.}(0)Songmuang, Giang, Bleuse, Hertog, Niquet,
  Dang, and Mariette]{Songmuang16}
Songmuang,~R.; Giang,~L. T.~T.; Bleuse,~J.; Hertog,~M.~D.; Niquet,~Y.~M.;
  Dang,~L.~S.; Mariette,~H. \emph{Nano Letters} \textbf{0}, \emph{0}, null,
  PMID: 27081785\relax
\mciteBstWouldAddEndPuncttrue
\mciteSetBstMidEndSepPunct{\mcitedefaultmidpunct}
{\mcitedefaultendpunct}{\mcitedefaultseppunct}\relax
\EndOfBibitem
\bibitem[Hocevar \latin{et~al.}(2013)Hocevar, Thanh~Giang, Songmuang, den
  Hertog, Besombes, Bleuse, Niquet, and Pelekanos]{Hocevar12}
Hocevar,~M.; Thanh~Giang,~L.~T.; Songmuang,~R.; den Hertog,~M.; Besombes,~L.;
  Bleuse,~J.; Niquet,~Y.-M.; Pelekanos,~N.~T. \emph{Applied Physics Letters}
  \textbf{2013}, \emph{102}, --\relax
\mciteBstWouldAddEndPuncttrue
\mciteSetBstMidEndSepPunct{\mcitedefaultmidpunct}
{\mcitedefaultendpunct}{\mcitedefaultseppunct}\relax
\EndOfBibitem
\bibitem[Heiss \latin{et~al.}(2013)Heiss, Fontana, Gustafsson, W\"ust, Magen,
  O'Regan, Luo, Ketterer, Conesa-Boj, Kuhlmann, Houel, Russo-Averchi, Morante,
  Cantoni, Marzari, Arbiol, Zunger, Warburton, and i~Morral]{Heiss13}
Heiss,~M. \latin{et~al.}  \emph{Nature Materials} \textbf{2013}, \emph{12},
  439\relax
\mciteBstWouldAddEndPuncttrue
\mciteSetBstMidEndSepPunct{\mcitedefaultmidpunct}
{\mcitedefaultendpunct}{\mcitedefaultseppunct}\relax
\EndOfBibitem
\bibitem[Mancini \latin{et~al.}(2014)Mancini, Fontana, Conesa-Boj, Blum,
  Vurpillot, Francaviglia, Russo-Averchi, Heiss, Arbiol, Morral, and
  Rigutti]{Mancini14}
Mancini,~L.; Fontana,~Y.; Conesa-Boj,~S.; Blum,~I.; Vurpillot,~F.;
  Francaviglia,~L.; Russo-Averchi,~E.; Heiss,~M.; Arbiol,~J.; Morral,~A. F.~i.;
  Rigutti,~L. \emph{Applied Physics Letters} \textbf{2014}, \emph{105}\relax
\mciteBstWouldAddEndPuncttrue
\mciteSetBstMidEndSepPunct{\mcitedefaultmidpunct}
{\mcitedefaultendpunct}{\mcitedefaultseppunct}\relax
\EndOfBibitem
\bibitem[Heiss \latin{et~al.}(2011)Heiss, Conesa-Boj, Ren, Tseng, Gali,
  Rudolph, Uccelli, Peir\'o, Morante, Schuh, Reiger, Kaxiras, Arbiol, and
  Fontcuberta~i Morral]{Heiss11}
Heiss,~M.; Conesa-Boj,~S.; Ren,~J.; Tseng,~H.-H.; Gali,~A.; Rudolph,~A.;
  Uccelli,~E.; Peir\'o,~F.; Morante,~J.~R.; Schuh,~D.; Reiger,~E.; Kaxiras,~E.;
  Arbiol,~J.; Fontcuberta~i Morral,~A. \emph{Phys. Rev. B} \textbf{2011},
  \emph{83}, 045303\relax
\mciteBstWouldAddEndPuncttrue
\mciteSetBstMidEndSepPunct{\mcitedefaultmidpunct}
{\mcitedefaultendpunct}{\mcitedefaultseppunct}\relax
\EndOfBibitem
\bibitem[Plissard \latin{et~al.}(2010)Plissard, Dick, Larrieu, Godey, Addad,
  Wallart, and Caroff]{Plissard10}
Plissard,~S.; Dick,~K.~A.; Larrieu,~G.; Godey,~S.; Addad,~A.; Wallart,~X.;
  Caroff,~P. \emph{Nanotechnology} \textbf{2010}, \emph{21}, 385602\relax
\mciteBstWouldAddEndPuncttrue
\mciteSetBstMidEndSepPunct{\mcitedefaultmidpunct}
{\mcitedefaultendpunct}{\mcitedefaultseppunct}\relax
\EndOfBibitem
\bibitem[Thelander \latin{et~al.}(2011)Thelander, Caroff, Plissard, Dey, and
  Dick]{Thelander11}
Thelander,~C.; Caroff,~P.; Plissard,~S.; Dey,~A.~W.; Dick,~K.~A. \emph{Nano
  Lett.} \textbf{2011}, \emph{11}, 2424--2429\relax
\mciteBstWouldAddEndPuncttrue
\mciteSetBstMidEndSepPunct{\mcitedefaultmidpunct}
{\mcitedefaultendpunct}{\mcitedefaultseppunct}\relax
\EndOfBibitem
\bibitem[de~la Mata \latin{et~al.}(2012)de~la Mata, Magen, Gazquez, Utama,
  Heiss, Lopatin, Furtmayr, Fernández-Rojas, Peng, Morante, Rurali, Eickhoff,
  i~Morral, Xiong, and Arbiol]{deLaMata12}
de~la Mata,~M.; Magen,~C.; Gazquez,~J.; Utama,~M. I.~B.; Heiss,~M.;
  Lopatin,~S.; Furtmayr,~F.; Fernández-Rojas,~C.~J.; Peng,~B.; Morante,~J.~R.;
  Rurali,~R.; Eickhoff,~M.; i~Morral,~A.~F.; Xiong,~Q.; Arbiol,~J. \emph{Nano
  Letters} \textbf{2012}, \emph{12}, 2579--2586, PMID: 22493937\relax
\mciteBstWouldAddEndPuncttrue
\mciteSetBstMidEndSepPunct{\mcitedefaultmidpunct}
{\mcitedefaultendpunct}{\mcitedefaultseppunct}\relax
\EndOfBibitem
\bibitem[Arbiol \latin{et~al.}(2009)Arbiol, Estradé, Prades, Cirera, Furtmayr,
  Stark, Laufer, Stutzmann, Eickhoff, Gass, Bleloch, Peiró, and
  Morante]{Arbiol09}
Arbiol,~J.; Estradé,~S.; Prades,~J.~D.; Cirera,~A.; Furtmayr,~F.; Stark,~C.;
  Laufer,~A.; Stutzmann,~M.; Eickhoff,~M.; Gass,~M.~H.; Bleloch,~A.~L.;
  Peiró,~F.; Morante,~J.~R. \emph{Nanotechnology} \textbf{2009}, \emph{20},
  145704\relax
\mciteBstWouldAddEndPuncttrue
\mciteSetBstMidEndSepPunct{\mcitedefaultmidpunct}
{\mcitedefaultendpunct}{\mcitedefaultseppunct}\relax
\EndOfBibitem
\bibitem[Shtrikman \latin{et~al.}(2009)Shtrikman, Popovitz-Biro, Kretinin,
  Houben, Heiblum, Bukala, Galicka, Buczko, and Kacman]{Shtrikman09}
Shtrikman,~H.; Popovitz-Biro,~R.; Kretinin,~A.; Houben,~L.; Heiblum,~M.;
  Bukala,~M.; Galicka,~M.; Buczko,~R.; Kacman,~P. \emph{Nano Letters}
  \textbf{2009}, \emph{9}, 1506--1510, PMID: 19253998\relax
\mciteBstWouldAddEndPuncttrue
\mciteSetBstMidEndSepPunct{\mcitedefaultmidpunct}
{\mcitedefaultendpunct}{\mcitedefaultseppunct}\relax
\EndOfBibitem
\bibitem[Shtrikman \latin{et~al.}(2009)Shtrikman, Popovitz-Biro, Kretinin, and
  Heiblum]{Shtrikman09a}
Shtrikman,~H.; Popovitz-Biro,~R.; Kretinin,~A.; Heiblum,~M. \emph{Nano Letters}
  \textbf{2009}, \emph{9}, 215--219\relax
\mciteBstWouldAddEndPuncttrue
\mciteSetBstMidEndSepPunct{\mcitedefaultmidpunct}
{\mcitedefaultendpunct}{\mcitedefaultseppunct}\relax
\EndOfBibitem
\bibitem[Plochocka \latin{et~al.}(2013)Plochocka, Mitioglu, Maude, Rikken,
  Granados~del Aguila, Christianen, Kacman, and Shtrikman]{Plochocka13}
Plochocka,~P.; Mitioglu,~A.~A.; Maude,~D.~K.; Rikken,~G. L. J.~A.; Granados~del
  Aguila,~A.; Christianen,~P. C.~M.; Kacman,~P.; Shtrikman,~H. \emph{Nano
  Letters} \textbf{2013}, \emph{13}, 2442--2447\relax
\mciteBstWouldAddEndPuncttrue
\mciteSetBstMidEndSepPunct{\mcitedefaultmidpunct}
{\mcitedefaultendpunct}{\mcitedefaultseppunct}\relax
\EndOfBibitem
\bibitem[Yuan \latin{et~al.}(2015)Yuan, Caroff, Wong-Leung, Fu, Tan, and
  Jagadish]{Yuan15}
Yuan,~X.; Caroff,~P.; Wong-Leung,~J.; Fu,~L.; Tan,~H.~H.; Jagadish,~C.
  \emph{Advanced Materials} \textbf{2015}, \emph{27}, 6096--6103\relax
\mciteBstWouldAddEndPuncttrue
\mciteSetBstMidEndSepPunct{\mcitedefaultmidpunct}
{\mcitedefaultendpunct}{\mcitedefaultseppunct}\relax
\EndOfBibitem
\bibitem[Chi \latin{et~al.}(2013)Chi, Chang, Hu, Yeh, Cronin, and
  Dapkus]{Chun13}
Chi,~C.-Y.; Chang,~C.-C.; Hu,~S.; Yeh,~T.-W.; Cronin,~S.~B.; Dapkus,~P.~D.
  \emph{Nano Letters} \textbf{2013}, \emph{13}, 2506--2515, PMID:
  23634790\relax
\mciteBstWouldAddEndPuncttrue
\mciteSetBstMidEndSepPunct{\mcitedefaultmidpunct}
{\mcitedefaultendpunct}{\mcitedefaultseppunct}\relax
\EndOfBibitem
\bibitem[Arab \latin{et~al.}(2015)Arab, Chi, Shi, Wang, Dapkus, Jackson, Smith,
  and Cronin]{Arab15}
Arab,~S.; Chi,~C.-Y.; Shi,~T.; Wang,~Y.; Dapkus,~D.~P.; Jackson,~H.~E.;
  Smith,~L.~M.; Cronin,~S.~B. \emph{ACS Nano} \textbf{2015}, \emph{9},
  1336--1340, PMID: 25565000\relax
\mciteBstWouldAddEndPuncttrue
\mciteSetBstMidEndSepPunct{\mcitedefaultmidpunct}
{\mcitedefaultendpunct}{\mcitedefaultseppunct}\relax
\EndOfBibitem
\bibitem[Chang \latin{et~al.}(2014)Chang, Chi, Chen, Huang, Arab, Qiu,
  Povinelli, Dapkus, and Cronin]{Chang2014}
Chang,~C.~C.; Chi,~C.~Y.; Chen,~C.~C.; Huang,~N.; Arab,~S.; Qiu,~J.;
  Povinelli,~M.~L.; Dapkus,~P.~D.; Cronin,~S.~B. \emph{Nano Research}
  \textbf{2014}, \emph{7}, 163--170\relax
\mciteBstWouldAddEndPuncttrue
\mciteSetBstMidEndSepPunct{\mcitedefaultmidpunct}
{\mcitedefaultendpunct}{\mcitedefaultseppunct}\relax
\EndOfBibitem
\bibitem[de~la Mata \latin{et~al.}(2016)de~la Mata, Leturcq, Plissard, Rolland,
  Magén, Arbiol, and Caroff]{delaMata15}
de~la Mata,~M.; Leturcq,~R.; Plissard,~S.~R.; Rolland,~C.; Magén,~C.;
  Arbiol,~J.; Caroff,~P. \emph{Nano Letters} \textbf{2016}, \emph{16},
  825--833, PMID: 26733426\relax
\mciteBstWouldAddEndPuncttrue
\mciteSetBstMidEndSepPunct{\mcitedefaultmidpunct}
{\mcitedefaultendpunct}{\mcitedefaultseppunct}\relax
\EndOfBibitem
\bibitem[Tutuncuoglu \latin{et~al.}(2015)Tutuncuoglu, de~la Mata, Deiana,
  Potts, Matteini, Arbiol, and Fontcuberta~i Morral]{Tutuncuoglu15}
Tutuncuoglu,~G.; de~la Mata,~M.; Deiana,~D.; Potts,~H.; Matteini,~F.;
  Arbiol,~J.; Fontcuberta~i Morral,~A. \emph{Nanoscale} \textbf{2015},
  \emph{7}, 19453--19460\relax
\mciteBstWouldAddEndPuncttrue
\mciteSetBstMidEndSepPunct{\mcitedefaultmidpunct}
{\mcitedefaultendpunct}{\mcitedefaultseppunct}\relax
\EndOfBibitem
\bibitem[Heiß \latin{et~al.}(2008)Heiß, Riedlberger, Spirkoska, Bichler,
  Abstreiter, and i~Morral]{Heiß08}
Heiß,~M.; Riedlberger,~E.; Spirkoska,~D.; Bichler,~M.; Abstreiter,~G.;
  i~Morral,~A.~F. \emph{Journal of Crystal Growth} \textbf{2008}, \emph{310},
  1049 -- 1056\relax
\mciteBstWouldAddEndPuncttrue
\mciteSetBstMidEndSepPunct{\mcitedefaultmidpunct}
{\mcitedefaultendpunct}{\mcitedefaultseppunct}\relax
\EndOfBibitem
\bibitem[Woolf \latin{et~al.}(1992)Woolf, Sobiesierski, Westwood, and
  Williams]{Woolf92}
Woolf,~D.~A.; Sobiesierski,~Z.; Westwood,~D.~I.; Williams,~R.~H. \emph{Journal
  of Applied Physics} \textbf{1992}, \emph{71}, 4908--4915\relax
\mciteBstWouldAddEndPuncttrue
\mciteSetBstMidEndSepPunct{\mcitedefaultmidpunct}
{\mcitedefaultendpunct}{\mcitedefaultseppunct}\relax
\EndOfBibitem
\bibitem[Heim and Hiesinger(1974)Heim, and Hiesinger]{Heim74}
Heim,~U.; Hiesinger,~P. \emph{physica status solidi (b)} \textbf{1974},
  \emph{66}, 461--470\relax
\mciteBstWouldAddEndPuncttrue
\mciteSetBstMidEndSepPunct{\mcitedefaultmidpunct}
{\mcitedefaultendpunct}{\mcitedefaultseppunct}\relax
\EndOfBibitem
\bibitem[Kunzel and Ploog(1980)Kunzel, and Ploog]{Kunzel80}
Kunzel,~H.; Ploog,~K. \emph{Applied Physics Letters} \textbf{1980}, \emph{37},
  416--418\relax
\mciteBstWouldAddEndPuncttrue
\mciteSetBstMidEndSepPunct{\mcitedefaultmidpunct}
{\mcitedefaultendpunct}{\mcitedefaultseppunct}\relax
\EndOfBibitem
\bibitem[Zhang \latin{et~al.}(2009)Zhang, Tateno, Sanada, Tawara, Gotoh, and
  Nakano]{Zhang09}
Zhang,~G.; Tateno,~K.; Sanada,~H.; Tawara,~T.; Gotoh,~H.; Nakano,~H.
  \emph{Applied Physics Letters} \textbf{2009}, \emph{95}\relax
\mciteBstWouldAddEndPuncttrue
\mciteSetBstMidEndSepPunct{\mcitedefaultmidpunct}
{\mcitedefaultendpunct}{\mcitedefaultseppunct}\relax
\EndOfBibitem
\bibitem[Ploog and Döhler(1983)Ploog, and Döhler]{Ploog83}
Ploog,~K.; Döhler,~G.~H. \emph{Advances in Physics} \textbf{1983}, \emph{32},
  285--359\relax
\mciteBstWouldAddEndPuncttrue
\mciteSetBstMidEndSepPunct{\mcitedefaultmidpunct}
{\mcitedefaultendpunct}{\mcitedefaultseppunct}\relax
\EndOfBibitem
\bibitem[van Weert \latin{et~al.}(2006)van Weert, Wunnicke, Roest, Eijkemans,
  Yu~Silov, Haverkort, 't~Hooft, and Bakkers]{Weert06}
van Weert,~M. H.~M.; Wunnicke,~O.; Roest,~A.~L.; Eijkemans,~T.~J.;
  Yu~Silov,~A.; Haverkort,~J. E.~M.; 't~Hooft,~G.~W.; Bakkers,~E. P. A.~M.
  \emph{Applied Physics Letters} \textbf{2006}, \emph{88}\relax
\mciteBstWouldAddEndPuncttrue
\mciteSetBstMidEndSepPunct{\mcitedefaultmidpunct}
{\mcitedefaultendpunct}{\mcitedefaultseppunct}\relax
\EndOfBibitem
\bibitem[Signorello \latin{et~al.}(2013)Signorello, Karg, Björk, Gotsmann, and
  Riel]{Signorell013}
Signorello,~G.; Karg,~S.; Björk,~M.~T.; Gotsmann,~B.; Riel,~H. \emph{Nano
  Letters} \textbf{2013}, \emph{13}, 917--924, PMID: 23237482\relax
\mciteBstWouldAddEndPuncttrue
\mciteSetBstMidEndSepPunct{\mcitedefaultmidpunct}
{\mcitedefaultendpunct}{\mcitedefaultseppunct}\relax
\EndOfBibitem
\bibitem[Oskooi \latin{et~al.}(2010)Oskooi, Roundy, Ibanescu, Bermel,
  Joannopoulos, and Johnson]{Oksooi10}
Oskooi,~A.~F.; Roundy,~D.; Ibanescu,~M.; Bermel,~P.; Joannopoulos,~J.~D.;
  Johnson,~S.~G. \emph{Computer Physics Communications} \textbf{2010},
  \emph{181}, 687--702\relax
\mciteBstWouldAddEndPuncttrue
\mciteSetBstMidEndSepPunct{\mcitedefaultmidpunct}
{\mcitedefaultendpunct}{\mcitedefaultseppunct}\relax
\EndOfBibitem
\bibitem[Adachi(1985)]{Adachi1985}
Adachi,~S. \emph{Journal of Applied Physics} \textbf{1985}, \emph{58},
  R1--R29\relax
\mciteBstWouldAddEndPuncttrue
\mciteSetBstMidEndSepPunct{\mcitedefaultmidpunct}
{\mcitedefaultendpunct}{\mcitedefaultseppunct}\relax
\EndOfBibitem
\end{mcitethebibliography}

\end{document}